\documentclass[prc,nofootinbib,showpacs,preprintnumbers,amsmath,amssymb]{revtex4}

\usepackage{graphicx}
\usepackage{dcolumn}
\usepackage{bm}

\def\ii{\'{\i }}

\usepackage{epsfig}

\def\menorsim{\smash{\mathop{<}\limits_{\raise3pt\hbox{$\sim$}}}}

\begin{document}

\title{Forward-Backward rapidity correlations at all rapidities}%

\author{P. Brogueira}
\email{pedro@fisica.ist.utl.pt}
\affiliation{Departamento de F\ii sica, IST, Av. Rovisco Pais, 1049-001 Lisboa, Portugal }
\author{J. Dias de Deus}
\email{jdd@fisica.ist.utl.pt}
\affiliation{CENTRA, Departamento de F\ii sica, IST, Av. Rovisco Pais, 1049-001 Lisboa, Portugal}
\author{Jos\'e Guilherme Milhano}
\email{gui@fisica.ist.utl.pt}
\affiliation{CENTRA, Departamento de F\ii sica, IST, Av. Rovisco Pais, 1049-001 Lisboa, Portugal}

\date{\today}%

\begin{abstract}
We discuss forward-bacward rapidity correlations in the general situation of asymmetrical collisions, asymmetric rapidity windows, higher rapidities and higher energy. We give predictions for RHIC and LHC.
\end{abstract}


\maketitle

The study of correlations and fluctuations can give an important contribution to the understanding of the behaviour of dense matter in high energy heavy ion collisions. See, for recent experimental developments [1,2,3,4] and for recent theoretical studies [5,6,7,8].

We consider forward-backward (FB) rapidity correlations in  the framework of a two-step scenario [9,6,7,8] with formation of longitudinal sources, as it happens in the glasma [9] or in the dual string model [10]. Even in the case of particle emission from the sources being fully uncorrelated, long distance E-B correlations are unavoidable due to fluctuations in the colour or the number of sources. In the situation of multiple elementary collisions, as it occurs in nucleus-nucleus collisions, and hadron-hadron collision, as well, such fluctuations are naturally expected [11,12].

The basic formalism for the development of the two-step scenario is as follows. In a high energy nucleus-nucleus, or hadron-nucleus, or hadron-hadron collision the interaction occurs via $\nu$ elementary interactions with formation of sources: longitudinal fields or strings. In the dual string model- which we shall use as reference, but nothing essential depends on that -- the number $N_s$ of formed strings is directly related to the number of elementary collisions:

\begin{eqnarray}
N_s = 2\nu \ .
\end{eqnarray}

\noindent Next, the strings emit particles as a local, short range correlated in rapidity, phenomenon. If $P(n_F , n_B)$ is the probability of emitting $n_F$ particles in the  backward one, we have

\begin{eqnarray}
P(n_F ,n_B) = \sum_{\nu} \varphi (\nu)\sum_{\{ n_{iF} ,n_{iB}\}} \prod_i p(n_{iF},n_{iB}) \delta (n_F -\sum_i n_{iF} )\delta (n_B -\sum_i n_{iB}) 
\end{eqnarray}

\noindent where $\varphi (\nu)$ is the probability of occurence of $\nu$ elementary collisions and $p(n_{iF},n_{iB})$ the local probability of $n_{iF}$ particles in the forward window and $n_{iB}$ particles in the backward one from the i-th string.

We make now the usual assumption that particle emission from the string is not long range correlated such that, if a, large enough, rapidity gap exists between the forward and the backward window, then

\begin{eqnarray}
p(n_{iF}, n_{iB}) = p(n_{iF})\ p(n_{iB})
\end{eqnarray}

\noindent and, for the average of $n_{iF} n_{iB}$, 

\begin{eqnarray}
\overline{n_{iF} n_{iB}} = \overline{n_{iF}} \ \overline{n_{iB}} \ .
\end{eqnarray}

However, local short range correlations in the forward (and backward) window exist -- due, for instance, to resonnances -- and characterized by the local variance

\begin{eqnarray}
d^2_{iF} \equiv \overline{n^2_{iF}} - \overline{n_{iF}}^2 \ .
\end{eqnarray}

If the strings are non-interacting and are all identical and symmetrical with respect to central rapidity -- which requires symmetrical AA collisions and flat rapidity distribution -- then in any rapidity window one has all the $N_s$ strings contributing which means that in the summations $\sum_{i}$ and in the products $\prod_i$, $i$ takes the values, see (1),

\begin{eqnarray}
i= 1,2,\dots , 2\nu \ .
\end{eqnarray}

However, in general, (6) is not correct, as the number of strings contributing in a given rapidity window is, in general smaller than $2\nu$. There are two reasons for that. First, strings may fuse and percolate such that their number decreases [13,14]. Second, some strings are smaller than others -- that is the reason why the particle density in rapidity, $dn/dy$, is not flat -- which implies an additional rapidity dependent reduction. These reduction factors are naturally functions of the window rapidities, $y_F$ and $y_B$, respectively.

Instead of the limit $2\nu$ in (6) we thus write for the number of contributing strings

\begin{eqnarray}
F: \ 2\nu \longrightarrow 2\nu r (y_F) \ , \\
B: \ 2\nu \longrightarrow 2\nu r (y_B) \ , 
\end{eqnarray}

\noindent where $r(y_F)$ and $r(y_B)$ are the reduction factors, with $0<r(y_F), r(y_B) \leq 1$. As one imediately sees, we do not have to make estimates on $r(y_F)$ and $r(y_B)$ has they appear via the measurable multiplicities $\left\langle n_F\right\rangle$ and $\left\langle n_B\right\rangle$.

With (3), (5), (7) and (8) in (2) we obtain:

\begin{eqnarray}
\left\langle n_F\right\rangle = 2 <\nu > r (y_F) \bar n_F \ , \\
\left\langle n_B\right\rangle = 2 <\nu > r (y_B) \bar n_B \ , \\
\nonumber \\
D^2_{FB} = \left\langle n_F n_B \right\rangle - \left\langle n_F\right\rangle \left\langle n_B \right\rangle= {\left\langle n_F\right\rangle \left\langle n_B \right\rangle \over K} \ ,
\end{eqnarray}

\noindent and

\begin{eqnarray}
D^2_{FF} = \left\langle n_F^2\right\rangle - \left\langle n_F\right\rangle^2 = {\left\langle n_F \right\rangle^2 \over K} + {d_F^2 \over \bar n_F} \left\langle n_F \right\rangle \ ,
\end{eqnarray}

\noindent where $1/K$ is the normalized fluctuation of the $\nu$ distribution:

\begin{eqnarray}
K\equiv {\left\langle \nu \right\rangle^2 \over \left\langle \nu^2 \right\rangle - \left\langle \nu \right\rangle^2} \ .
\end{eqnarray}

Usually, one defines the $F-B$ correlation parameter $b$ as

\begin{eqnarray}
b\equiv {D^2_{FB} \over D^2_{FF}} \ ,
\end{eqnarray}

\noindent such that, from (9), (10), (11), (12) and (14), we obtain 

\begin{eqnarray}
b= {\left\langle n_B \right\rangle / \left\langle n_F \right\rangle \over 1+ {K\over \left\langle n_F \right\rangle} {d^2_F \over \bar n_F}} \ . 
\end{eqnarray}

\noindent With symmetrical situation, $AA$ collisions, and $y_B=-y_F, \left\langle n_B \right\rangle = \left\langle n_F\right\rangle$, we obtain the well known result [15,7]:

\begin{eqnarray}
b= {1\over 1+ {k\over \left\langle n_F\right\rangle} {d^2_F \over \bar n_F}} \ .
\end{eqnarray}

\noindent It is clear from (15) that there exists the possibility of $b$ being larger than 1. This is particularly true if one neglects short ramge fluctuations: $d^2_F \equiv 0$, and

\begin{eqnarray}
b\simeq \left\langle n_B\right\rangle / \left\langle n_F\right\rangle \ .
\end{eqnarray}

In general it is assumed that local emission follows a Poisson distribution, such that $d^2_F = \bar n_F$ and

\begin{eqnarray}
b={\left\langle n_B\right\rangle / \left\langle n_F\right\rangle \over 1+K/\left\langle n_F\right\rangle } \ .
\end{eqnarray}

\noindent One should keep in mind that (16) and (17) are quite general, in the sense that they do not require symmetric $AA$ collisions or symmetric windows.

The framework of the two-step scenario with long range correlations due to extended sources in rapidity and local short range correlations, is also present in the colour Glass Condensate (CGC) approach [6,8]. The result obtained has the expected structure of (15) or (18): $b=A[1+B]^{-1}$.

Let us test the validity of (18) for $b$ and (11) and (12) for $D^2_{FB}$ and $D^2_{FF}$, respectively, in the symmetrical situation, $AA$ collisions and $y_B = -y_F (\left\langle n_F\right\rangle = \left\langle n_B\right\rangle)$. The STAR collaboration at RHIC has presented new results for $D^2_{FB}$ and $D^2_{FF}$ in central collisions at mid rapidity, making use of windows of width $\delta \eta =0.2$ and a distance between their centres of $\Delta \eta (0.2 < \Delta \eta < 1.8)$ [3]. In (11) and (12) $K$ has fixed  at the value of 88\footnote{As the STAR value for $D^2_{FB}$ went up by a factor of around 4 [3], the value of $K$ used in [7], goes down by the same amount.}.

\begin{figure}[t]
\begin{center}
\includegraphics[width=9cm]{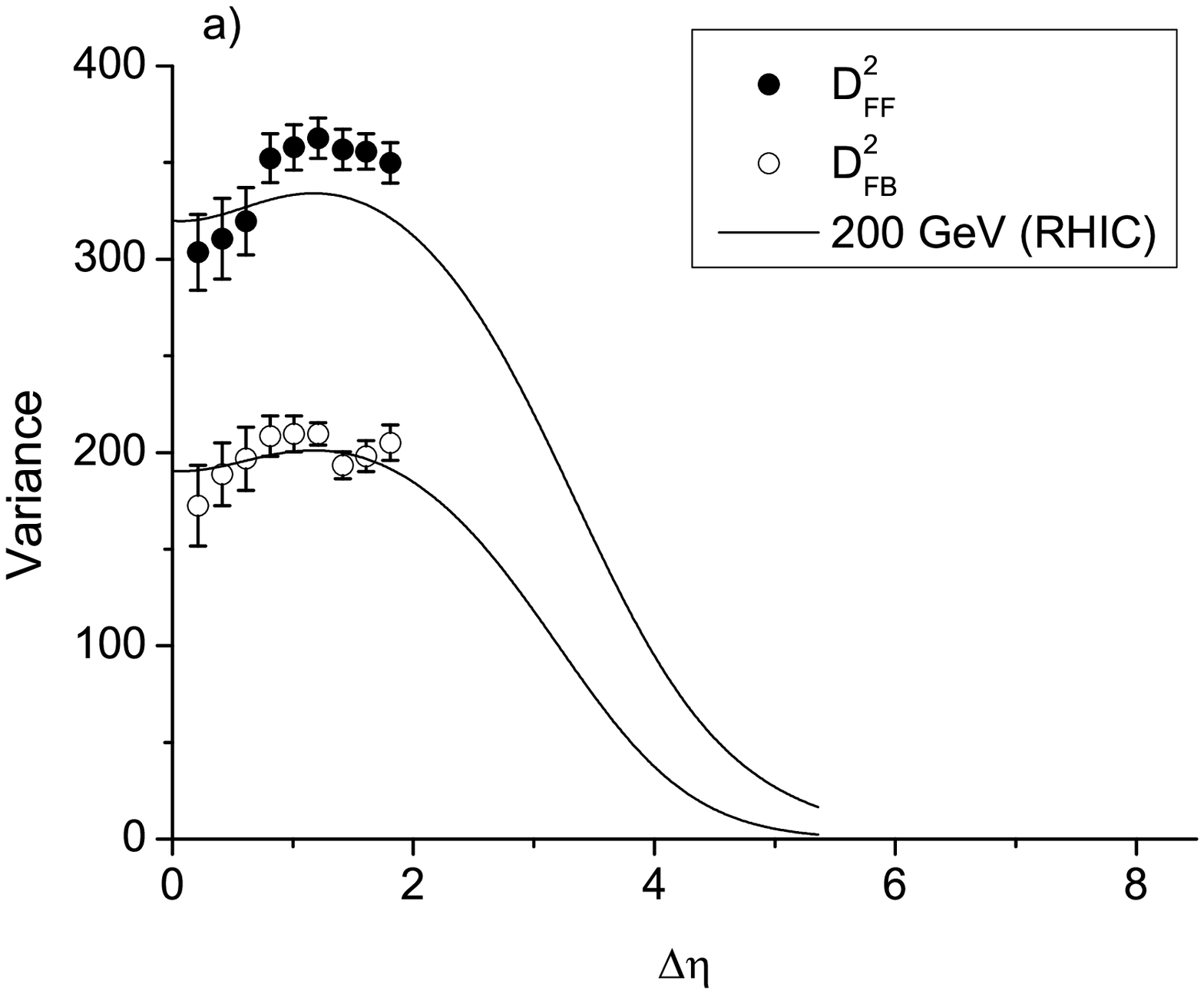}\includegraphics[width=9cm]{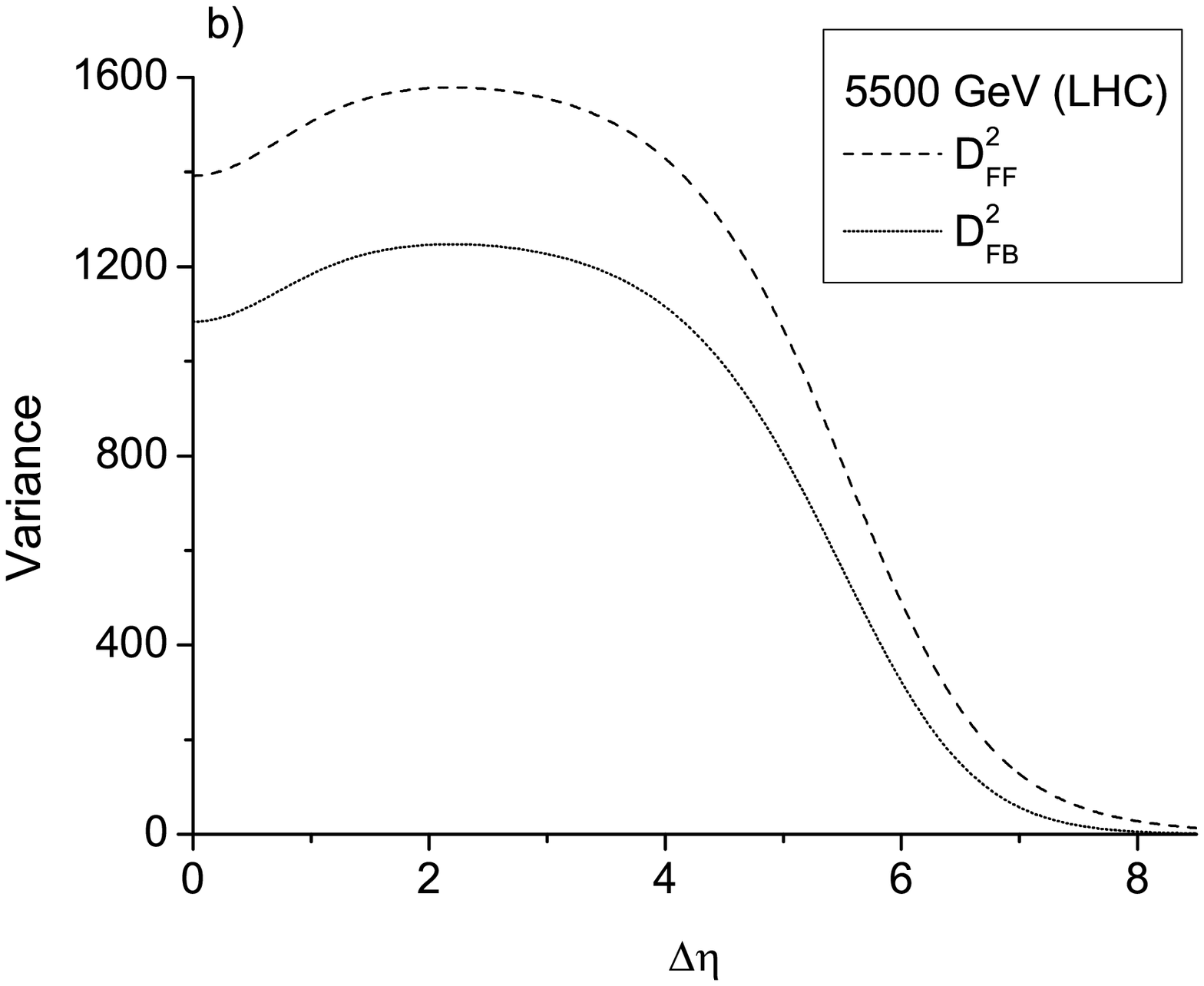}      
\end{center}
\caption{a) STAR central data, 0-0.6\%, $Au-Au$ data at RHIC, and our curves (full lines) for $D^2_{FB}$ (11) and $D^2_{FB}$ (12). b)Predictions for LHC, $Pb-Pb$ central (dashed lines).}
\end{figure}

In Fig. 1a) we compare (11) and (12) with mid rapidity data for central $Au-Au$ at $\sqrt{s} =200$GeV and show our expectation in forward rapidity region. In Fig. 1b) we show  our prediction for $Pb-Pb$ at LHC, $\sqrt{s} = 5500$GeV. In Fig. 2 we compare $b$, (18) with RHIC data and we show as well our predictions for LHC. Note that our prediction qualitatively agrees with the CGC model [6]. In our plots we have used for highest centrality particle density the parameterization of [16], 

\begin{eqnarray}
{1\over A} {dn \over dy} = {e^{\lambda Y} \over e^{\eta-(1-\alpha)Y\over \delta} +1} \ , 
\end{eqnarray}

\noindent where $Y$ is the beam rapidity. The parameters in (19) where ajusted to fit the PHOBOS/RHIC $dn/d \eta$ data at $\sqrt{s} =200$GeV  [17]: $\lambda = 0.26,\ \alpha =0.31,\ \delta=0.75$.

\begin{figure}[t]
\begin{center}
\includegraphics[width=11cm]{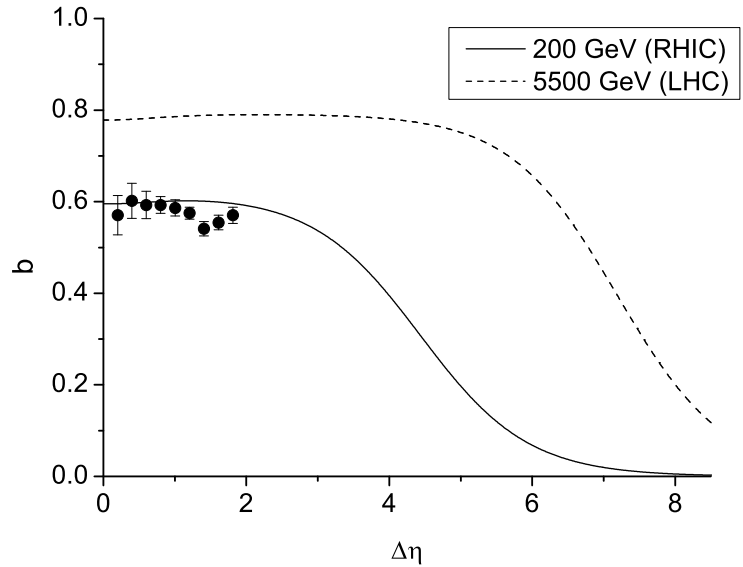}      
\end{center}
\caption{STAR data and our curve (full line) for $b$, (18). Predictions for LHC, $Pb-Pb$ central (dashed lines).}
\end{figure}

Regarding the parameter $K$, (13), which is defined at partonic/string level, we would like to mention that, in the case of $AA$ collisions, $K$ is related to a similar parameter $K_n$, defined at nucleon level and which can be estimated in a Glauber calculation [18]. As $K_n$ depends mostly on the nucleus-nucleon cross-section and $\sigma_{nn}$ slowly varies with energy, one expects $K_n$ to be fairly independent of energy. On the other hand, repeating the arguments of [12], which essentially says that the normalized fluctuation at the parton level should be larger than the same quantity at nucleon level, we obtain:

\begin{eqnarray}
K\menorsim K_n \simeq const.
\end{eqnarray}

\noindent The first relation in (20) is satisfied [18].

Having in mind (20) and the fact that $K$ is a KNO (Koba, Nielsen, Olesen) invariant quantity [19], we have assumed.

\begin{eqnarray}
K = const. ,
\end{eqnarray}

\noindent independent of energy, for our predictions at LHC, in Figs.1 and 2. Note that $K$, similarly to $K_n$, is expected to increase with the number of participating nucleons.

In the general case of asymmetrical $AB$ collisions and asymmetrical windows one can use the parameterizations of [20].

There is a simple way of testing the model for forward-backward correlations in the general case (asymmetric distributions and asymmetric windows). Let us fix $y_B$, the centre of the backward window, and move $y_F$, the centre of the forward window, with $y_F > y_B$. We can rewrite (18) in the form

\begin{eqnarray}
b = {x\over 1+K' x}
\end{eqnarray}

\noindent with

\begin{eqnarray}
x\equiv \left\langle n_B\right\rangle / \left\langle n_F\right\rangle \ ,
\end{eqnarray}

\begin{figure}
\includegraphics[width=11cm]{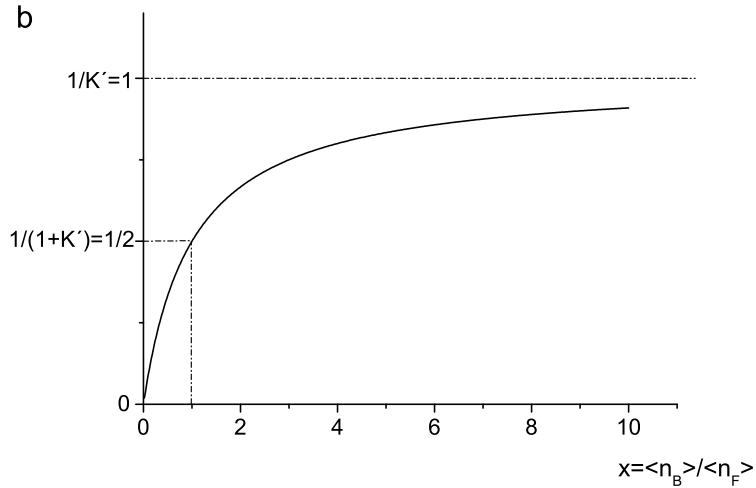}      
\caption{The correlation parameter $b$ as a function of $x\equiv \left\langle n_B\right\rangle / \left\langle n_F\right\rangle \ $. The function (21) is universal, only depending on the value of 
$K' \equiv K /\left\langle n_B\right\rangle \ $. In the figure $K'=1$.}
\end{figure}

\noindent $0<x<\infty$, and

\begin{eqnarray}
K' \equiv K /\left\langle n_B\right\rangle \ .
\end{eqnarray}

\noindent Independentely of the shape of the rapidity density distribution, of the centrality, of the energy, of the rapidity size of the windows, the functional form of the dependence of $b$ on $x$ is given by (22). Of course $K'$ is a number that varies with centrality (number of participating nucleons), energy and window size. However, the shape of the $b$ dependence on $x$ is always given by (22). In Fig. 3 we show the dependence of $b$ on $x$ (in the case of $K'\equiv 1$).

Finally, instead of studying forward-backward correlations, backward-forward correlations can as well be studied. One has simply to make the changes $B\to F$ and $F \to B$ in our formulae.

\bigskip
\bigskip

{\it Acknowledgments}

\bigskip

We would like to thank Nestor Armesto and Carlos Pajares for discussions.

\bigskip
\bigskip

{\it References}
\bigskip

\begin{enumerate}
\item B.B. Back et al. (PHOBOS Coll.), Phys. Rev. C74 (2006) 011901.
\item M. Rybczynski et al. (NA49 Coll.), nucl-ex/0409009 (2004). 
\item T.J. Tarnowsky (STAR Coll.), nucl-ex/0606018, Proc. 22nd Winter Workshop on Nuclear Dynamics (2006); B.K. Srivastava (STAR Coll.), nucl-ex/0702054 (2007).
\item J.T. Mitchell (PHENIX Coll.), nucl-ex/0511033 (2005).
\item S. Haussler, M. Abdel-Aziz, M. Bleicher, nucl-th/0608021.
\item N. Armesto, L. McLerran and C. Pajares, Nucl. Phys. A781 (2007) 201.
\item P. Brogueira, J. Dias de Deus, hep-ph/0611329 v2 (2006), Phys. Lett. B 653, 202 (2007).
\item N. Armesto, M.A. Braun, C. Pajares, hep-ph/0702216 v2 (2007).
\item A. Kovner, L.D. McLerran and H. Weigert, Phys. Rev. D52 (1995) 6231; T. Lappi and L.D. McLerran, Nucl. Phys. A772 (2006) 200.
\item N.S. Amelin, N. Armesto, M.A. Braun, E.G. Ferreiro, C. Pajares, Phys. Rev. Lett. 73 (1994) 2813.
\item A. Capella, A. Krzywicki, Phys. Rev. D18 (1978) 4120; A. Capella, J. Tran Thanh Van, Phys. Rev. D29 (1984) 2512.
\item J. Dias de Deus, C. Pajares, C. Salgado, Phys. Lett. B407 (1997) 335.
\item T.S. Bico, H.B. Nielsen, J. Knoll, Nucl Phys. B245 (1984) 449.
\item M.A. Braun, F. Del Moral, C. Pajares, Phys. Rev. C65 (2002) 024907.
\item M.A. Braun, C. Pajares, V.V. Vechernin, Phys. Lett. B498 (2000) 54.
\item P. Brogueira, J. Dias de Deus and C. Pajares, hep-ph/0605148 (2006),  Phys. Rev. C75:054908 (2007).
\item B. B. Back et al. (PHOBOS Collaboration) Nucl. Phys. A 757, 28 (2005).
\item V.V. Vechernin, hep-ph/0702141 (2007); G. Feofilov and A. Ivanov, Journ. Phys. C5 (2005) 230.
\item Z. Koba, H.B. Nielsen and P. Olesen, Nucl. Phys. B40, 317 (1972).
\item J. Dias de Deus and J.G. Milhano, hep-ph/0701215 (2007), to be published in Nucl. Phys. A.
\end{enumerate}

\end{document}